\documentclass[aps,amsfonts,prl,nofootinbib,tightenlines,preprint,superscriptaddress,showpacs]{revtex4}

\usepackage{bm}
\usepackage{epsfig}
\usepackage{subfigure}
\usepackage{amsmath}

\def\mit{\affiliation{Center for Theoretical Physics, Laboratory for
Nuclear Science, and Department of Physics,
Massachusetts Institute of Technology,
Cambridge, MA 02139}}
\def\midd{\affiliation{Department of Physics, Middlebury College,
Middlebury, VT  05753}}
\begin{document}

\title{Emergence of Oscillons in an Expanding Background}

\author{E.~Farhi}
\email{farhi@mit.edu}
\mit

\author{N.~Graham}
\email{ngraham@middlebury.edu}
\midd
\mit

\author{A.~H.~Guth}
\email{guth@ctp.mit.edu}
\mit

\author{N.~Iqbal}
\email{niqbal@mit.edu}
\mit

\author{R.~R.~Rosales}
\email{rrr@mit.edu}
\affiliation{
Department of Mathematics,
Massachusetts Institute of Technology,
Cambridge, MA 02139\\} 

\author{N.~Stamatopoulos}
\email{nstamato@middlebury.edu}
\altaffiliation[Current address: ]{Department of Physics and
Astronomy, Dartmouth College, Hanover, NH 03755.}
\midd

\preprint{MIT-CTP 3924}
\pacs{11.27.+d 05.70.Ln 98.80.Cq}

\begin{abstract}
We consider a (1+1) dimensional scalar field theory that supports
oscillons, which are localized, oscillatory, stable solutions to
nonlinear equations of motion.  We study this theory in an expanding
background and show that oscillons now lose energy, but at a rate
that is exponentially small when the expansion rate is slow.  We also
show numerically that a universe that starts with
(almost) thermal initial conditions will cool to a final state
where a significant fraction of the energy of the universe --- on the
order of 50\% --- is stored in oscillons.  If this phenomenon persists
in realistic models, oscillons may have cosmological consequences.
\end{abstract}
\maketitle

\section{Introduction}

A wide range of nonlinear field theories have been found to contain 
long lived, localized, oscillatory solutions to their equations of
motion, known as oscillons or breathers.  In the best-known examples
\cite{DHN,ColemanQ}, conserved charges guarantee the existence of exact
periodic solutions.  However, in many cases where such arguments are not
available, objects with similar properties have also been observed
\cite{Campbell,Bogolyubsky,Gleiser,2d,Honda,Wojtek,abelianhiggs,Forgacs,Gleiserd,oscillon,oscsm,Gleiserphase,Gleiserinfl,Kolb,GleiserU1,hindmarsh}.

In this work we consider analytically the evolution of an oscillon in
an inflating background.  We find that the oscillon is no longer
localized and contains a ``tail'' that slowly leaks energy
away.  However this decay rate is exponentially suppressed, so we
still expect there to exist long-lived objects.  This result is in
qualitative agreement with numerical work in \cite{oscex}, which
studied the long-term evolution of a single oscillon in an expanding
universe background and found that oscillons remain stable for an
exponentially long time provided that the horizon is far larger than
the width of an oscillon.

Once a model has been shown to contain oscillons, it is natural to ask
how easy it is for these coherent objects to form from generic
initial conditions.  Ref. \cite{Gleiserphase} showed that oscillons can
emerge from a rapid ``quench'' in which the background potential is
suddenly changed, throwing the system far out of equilibrium.  Here we
consider the opposite situation:  We begin with an (almost) thermal
distribution at high temperature and gradually cool the system by
coupling it to an expanding background.  This setup is suggestive of
a situation that could arise in the early universe, for example as the
universe cools after reheating or a phase transition.  If stable
oscillons formed in these situations, they might have cosmological
consequences: e.g., see \cite{Gleiserinfl}.  Although it did not
consider oscillons, \cite{Rajantie} found significant non-thermal
effects in electroweak baryon number violating processes.
Ref. \cite{riotto} studied a case similar to ours, but involving a
different type of oscillon that is stable only if its amplitude
exceeds a certain critical value.  This situation is quite different
from our model, in which there is no such threshold amplitude.

\begin{figure}[htbp]
\centerline{
\mbox{\includegraphics[width=0.71\linewidth]{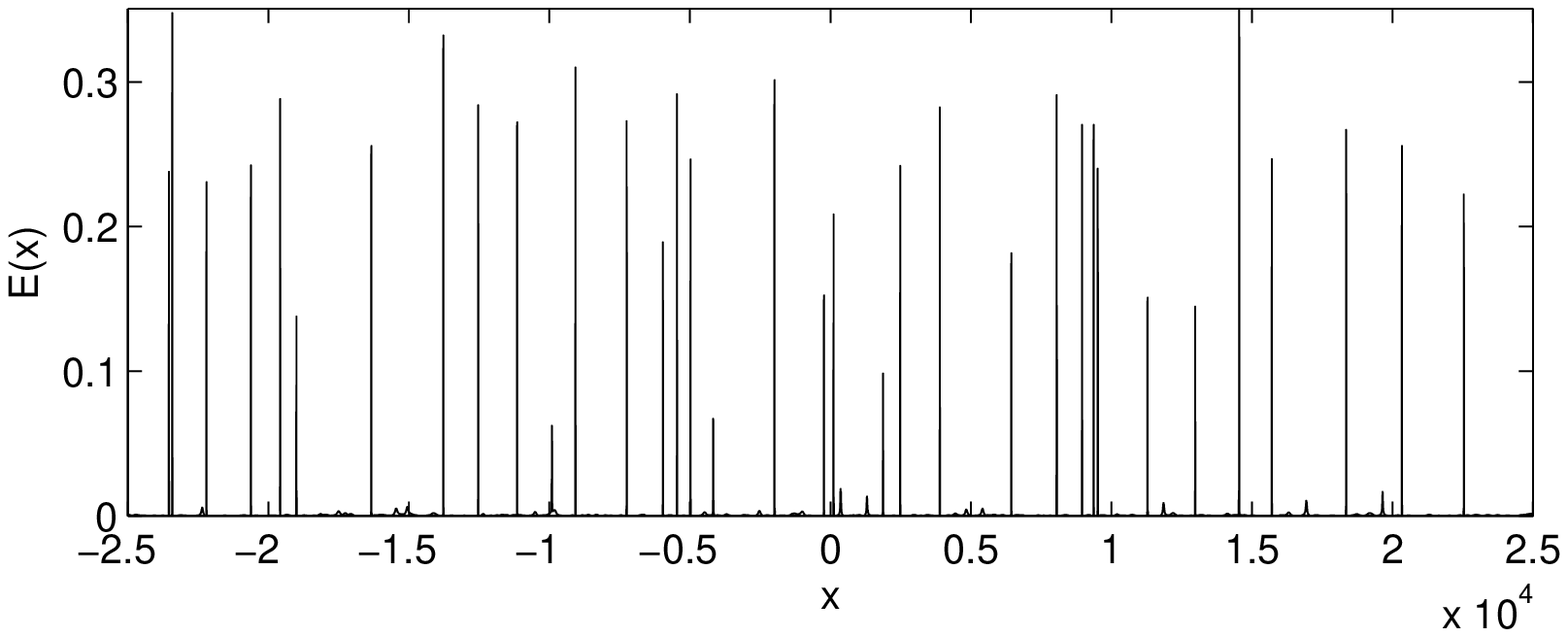}}
\mbox{\includegraphics[width=0.27\linewidth]{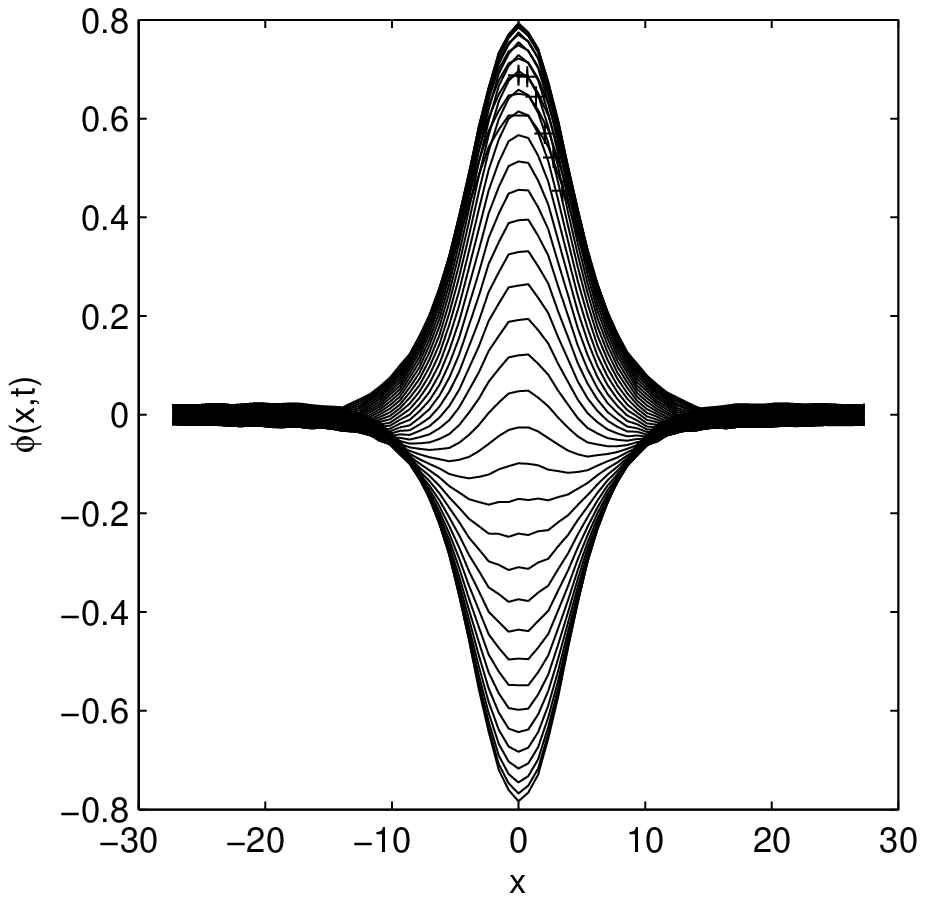}}
}
\caption{
(a) Energy density as a function of position for a typical final
state: $m = 1, \hbar = 0.5$.  (b) Snapshots of field profile
$\phi(x,t)$ for a  typical oscillon, taken at intervals of $\Delta t = 0.1.$
}
\label{fig3}
\end{figure}

For numerical convenience, we work with a single scalar field in one
dimension, though we have seen qualitatively similar results in
simulations of the two-dimensional scalar model of \cite{2d} in an
expanding background.  We start at a temperature for which the universe
is dominated by radiation and allow the universe to expand until it
contains only  a cold, pressureless dust of both oscillons and fundamental
excitations of the field.  A late-time snapshot of the energy density
of such a configuration  as a function of position is shown in Figure
\ref{fig3}, where the sharp spikes are oscillons.  We find that a
sizeable fraction of the energy of this final state --- of order 50\%
or more --- is stored in oscillons.  This result persists even for
very small coupling constants, where quantum effects are small and do
not affect our classical field theory analysis.

\section{Model}

We consider a toy model consisting of a massive scalar field
$\phi(x,t)$ in a one-dimensional expanding background.  In terms of
the comoving coordinate $x$, the model is described by the Lagrangian
\begin{equation}
{\cal L} = \frac{1}{2} \int \left[\dot \phi^2 - 
\frac{1}{a(t)^2} (\phi')^2 - m^2 \left(\phi^2
- \frac{g}{2} \phi^4 + \frac{g^2}{3} \phi^6 \right) \right] a(t) \,
dx \,
\label{lag}
\end{equation}
where an overdot denotes a derivative with respect to $t$ and a prime
denotes a derivative with respect to $x$.  Note that this potential
has a unique minimum at $\phi=0$, with no other extrema or inflection
points.  It does not support static solitons, but does support oscillons.
This Lagrangian leads to the equation of motion
\begin{equation}
\ddot \phi + \frac{\dot a(t)}{a(t)} \dot \phi = 
\frac{\phi''}{a(t)^2} - m^2 \left( \phi - g\phi^3 + g^2\phi^5 \right).
\label{eom}
\end{equation}
For simplicity we take the expansion rate $H=\dot a(t)/a(t)$ to be
constant.  Although we hope that our model can shed light on the
reheating phase of inflation, during which $H$ would be rapidly
decreasing, for our purposes it is only important that the expansion
rate be slow compared to the typical time scales of the oscillon.

All known oscillon solutions have been found for massive fields, with
frequency of oscillation below the threshold for fundamental
excitations, so we have chosen a massive field.  The sign of the
$\phi^4$ term, the leading nonlinearity, is crucial:  it has been
chosen so that to leading order this term decreases the frequency of
small oscillations.  With the other sign we do not see any oscillons
form.  Finally, the $\phi^6$ term exists to eliminate
instabilities at large field values introduced by the choice of sign
of the $\phi^4$ term.  Because the oscillons involve only moderate
excitations of the field, this term does not play an important role in
their dynamics.

We have scaled the self-interaction terms by a small
parameter $g$.  The exact meaning of this parameter is somewhat subtle,
since by defining a new field variable $\bar{\phi} \equiv \sqrt{g}\phi$
we can shift $g$ completely outside the Lagrangian (\ref{lag}), which means
that it no longer appears in the classical equations of motion.  The full
quantum theory is still sensitive to the value of $g$, but only in the
combination $\hbar g$; therefore a classical approximation involving small
$\hbar$ is equivalent to one involving small $g$, and the results of
our classical analysis are valid provided that $\hbar g$ is
small.\footnote{Also, throughout this paper the quantity we call $m$
has dimensions of [length]$^{-1}$, not energy, and is thus really the
mass of a single elementary quantum divided by $\hbar$.}

\section{Small-amplitude analysis of oscillons}

To describe the oscillons we expect to emerge from the thermal
background in our simulation, we use a small-amplitude analysis,
following \cite{smallamp}.  In particular (e.g. see \cite{oscex}) in a
static background the equation of motion  \eqref{eom} supports
solutions that are localized in space and periodic in time.  These
oscillons can be expanded in a small parameter $0 < \epsilon \ll 1$.
The amplitude of the oscillations scales with $\epsilon$, the
spatial width of the oscillon is proportional to $\epsilon^{-1}$, and
the oscillation frequency is given by $\omega =
m\,\sqrt{1-\epsilon^2}$.  Numerical calculations indicate that there
is an upper bound on how large $\epsilon$ can be for a stable
solution: $0 < \epsilon <\epsilon_c < 1$, where $\epsilon_c \approx 0.25$.

Here we extend these results to an inflationary background.
A sufficiently fast expansion will result in a horizon that is
comparable to the minimum oscillon width (proportional to
$\epsilon_c^{-1}$), preventing oscillons from ever forming.  If the
expansion rate is slow, long lived oscillons can occur in the range
$\sqrt{H/m} \ll \epsilon \ll 1$, as shown below.  In this regime the
oscillon radiates energy away (in the form of scalar field waves) at a
rate that is exponentially small in the dimensionless ratio
$m\epsilon^2/H$.  If this ratio becomes sufficiently large, numerical
calculations show that the oscillon is not affected by the expansion
in any way that we can detect.

\subsection{Separation of scales}

We work in static patch coordinates on de Sitter space (e.g., see
\cite{desitter}) where the metric takes the form
\begin{equation}
ds^2 = -(1-X^2 H^2)\;dT^2 + (1-X^2 H^2)^{-1}\;dX^2 \, .
\end{equation}
These coordinates are valid for $|XH| < 1$.  In these coordinates the
equation of motion (\ref{eom}) becomes
\begin{equation}
\frac{1}{(1-X^2H^2)}\;\ddot{\phi} + 2XH^2\phi' - (1-X^2H^2)\phi'' =
-m^2\left(\phi - g \phi^3 + g^2 \phi^5 \right) \,.
\label{eq:desitter_eom}
\end{equation}
Here an overdot indicates a derivative with respect to $T$ and a prime
indicates a derivative with respect to $X$.  Now we follow
\cite{oscex} and change to variables $\chi = mX\epsilon$ and $\tau =
mT\sqrt{1-\epsilon^2}$, where $0 < \epsilon \ll 1$ is the small
parameter mentioned above.  Note that since the horizon distance is
$1/H$ and the oscillon width is $1/m{\epsilon}$, to obtain a stable
oscillon solution we expect $H \ll m\epsilon$.  We therefore let
$H = m\bar{H}\epsilon^2$, where $\bar{H}$ is taken to be a
small dimensionless number.  The power of $\epsilon$ in this
expression is important:  If instead we take $H$ to be
$\mathcal{O}(\epsilon)$ then no oscillon solution is possible,
while if we take $H$ to be $\mathcal{O}(\epsilon^3)$ then (to the
order we are working) the oscillon does not feel the expansion.

In terms of $\chi$ and $\tau$ the equation of motion is 
\begin{equation}
\frac{(1-\epsilon^2)}{1-\chi^2\bar{H}\epsilon^{2}}(\phi)_{\tau\tau} +
2\chi\bar{H}^2\epsilon^{4}(\phi)_{\chi} -
(1-\chi^2\bar{H}\epsilon^{2})\epsilon^2(\phi)_{\chi\chi}
= -\left(\phi - g \phi^3 + g^2 \phi^5 \right) \, .
\label{eq:eom_curved}
\end{equation}
We expand $\phi$ in powers\footnote{Note that the expansion
\eqref{eq:eps_exp} involves only odd powers of $\epsilon$.  This is
consistent with \eqref{eq:eom_curved} because this equation is odd in
$\phi$, and depends on $\epsilon$ via $\epsilon^2$ only.} of
$\epsilon$ as
\begin{equation}
\phi(\chi, \tau) = \frac{1}{\sqrt{g}} \left[
\epsilon \phi_1(\chi, \tau) + 
\epsilon^3 \phi_3(\chi, \tau)  + \epsilon^5 \phi_5(\chi, \tau) +
\ldots \right]\, ,
\label{eq:eps_exp}
\end{equation} 
and seek oscillons, solutions that are periodic (with period $2\pi$) in
$\tau$ and localized in space.  In fact, as we will show
below, these solutions are neither strictly periodic
nor localized, but these departures are exponentially small
and occur over time scales much longer than the one given by $\tau$.

We now substitute \eqref{eq:eps_exp} into the governing equation
\eqref{eq:eom_curved} and solve it order by order in $\epsilon$.  At
$\mathcal{O}(\epsilon)$,
\begin{equation}
(\phi_1)_{\tau\tau} + \phi_1 = 0 \,.
\end{equation}
Thus $\phi_1 = f(\chi)e^{-i\tau} + f^*(\chi)e^{i\tau}$, where the
profile $f(\chi)$ remains to be determined.  We can find this profile
by considering the ${\cal O}(\epsilon^3)$ equation
\begin{equation}
(\phi_3)_{\tau\tau} + \phi_3 = (1 - \chi^2\bar{H}^2)(\phi_1)_{\tau\tau}
 + (\phi_1)_{\chi\chi} + \phi_1^3 \, .
\label{eq:nis2}
\end{equation}
The key point here is that the only way that $\phi_3$ can be periodic
in $\tau$ is if the forcing terms on the right-hand side of the
equation are orthogonal to $e^{\pm i\tau}$, since otherwise $\phi_3$
would have a component that grows linearly in $\tau$.  Therefore we
set the Fourier coefficients of $e^{\pm i \tau}$ to $0$, and obtain a
self-contained ordinary differential equation for the oscillon
profile:
\begin{equation}
\frac{d^2f}{d\chi^2} + \left(\chi^2\bar{H}^2 - 1\right)f(\chi) +
3f(\chi)^2f^*(\chi) = 0 \,.
\label{eq:osc_profile}
\end{equation} 
For a ``perfect'' oscillon, a localized (exponentially decaying as
$\chi \rightarrow \pm\infty$) spatial profile is needed.  This,
however, is not quite possible.  The behavior of the above equation
changes at $\chi \approx \pm 1/\bar{H}$: for small $|\chi|$ the
equation is essentially the same as the equation for a flat-space
oscillon, while for large $|\chi|$ the $\chi^2\bar{H}^2$ term
dominates, giving oscillatory behavior that causes the oscillon to
radiate (a small amount of) energy away.  Note that this
change in behavior occurs well before the horizon, which is at $\chi =
\pm 1/(\epsilon\bar{H})$.

\subsection{Asymptotic Behavior}
First we examine the regime where $|\chi| \ll 1/\bar{H}$.  Here the
term arising from the expansion of the universe is negligible and
the equation reduces to the case of a static background.  Therefore,
we can take
\begin{equation}
f(\chi) \approx \frac{1}{2}\sqrt{\frac{8}{3}}\,\mathrm{sech}(\chi)
\quad \mbox{for} \quad |\chi| \ll \frac{1}{\bar{H}} \,.
\label{eq:nonlinearsoln}
\end{equation}
Note that since $0 < \bar{H} \ll 1$, the oscillon behaves like
$\sqrt{\frac{8}{3}}e^{-\chi}$ in the region $1 \ll \chi \ll
1/\bar{H}$.  In particular, while its amplitude does not quite vanish
(since we cannot take $|\chi| \to \infty$ in the equation
above), it fails to be fully localized only because of an
exponentially small tail, as is shown below.

We now relax the upper bound and consider the region where $\chi \gg
1$ (an entirely similar calculation can be done where $\chi \ll
-1$).  From the reasoning above, we see that on the left-hand side of
this region, the field is exponentially small.  Thus, provided that the
field does not grow too much as $\chi$ increases in this region, we
are justified in neglecting the nonlinear term in the equation,
though we must now take into account the effects of the
expansion.  Thus we obtain
\begin{equation}
\frac{d^2f}{d\chi^2} + f\left(\chi^2\bar{H}^2 - 1\right) = 0 \,.
\end{equation}
This equation can be put into a more familiar form by defining a new
coordinate $y = \chi\bar{H}$.  Then we obtain
\begin{equation}
-\bar{H}^2\frac{d^2f}{dy^2} - y^2f = -f \,.
\end{equation}
This is exactly the Schr\"odinger equation for the wave function of a
particle in an upside-down harmonic oscillator with energy $E = -1$,
with the dimensionless Hubble constant $\bar{H}$ playing the same role
as Planck's constant $\hbar$ in the analogous quantum mechanics
problem.  At small $y$ the particle is in the ``classically forbidden''
region of the potential and the wavefunction is exponentially
suppressed, corresponding to the exponential tail of the oscillon.  For
any nonzero value of $\bar{H}$, we eventually enter a ``classically
allowed'' region and the wavefunction becomes oscillatory, which for
our classical field profile indicates an  outgoing wave that carries
energy away.  Therefore this situation looks exactly like a quantum
mechanics tunneling problem, and since we are interested in small
$\bar{H}$ we can use a semiclassical approximation to solve
it.\footnote{Of course, exact formal solutions for the wavefunction of
an upside-down oscillator exist in terms of Hermite functions.
However, extracting the asymptotic behavior from these special
functions is somewhat messy; for our purposes we can obtain equivalent
results simply using WKB.}

Choosing an outgoing wave boundary condition as $y\rightarrow\infty$,
the standard WKB connection formulae \cite{gottfried} give us the
following relation between the wavefunctions on either end of the
turning point at $y = 1$:
\begin{equation}
\frac{A}{\left(y^2-1\right)^{1/4}}\exp
\left[\frac{i}{\bar{H}}\int^y_1\sqrt{(y'^2-1)} dy' + \frac{i\pi}{4}\right]
\rightarrow
\frac{A}{\left(1-y^2\right)^{1/4}}\exp
\left[\frac{1}{\bar{H}}\int_y^1\sqrt{(1-y'^2)} dy'\right] \, , \nonumber
\end{equation}
where the left-hand side is valid for $y \gg 1$, the right-hand side
is valid for $y \ll 1$, and $A$ is an overall constant.

Performing the integrals, keeping only the leading dependence, and
replacing $y$ with $\chi\bar{H}$, we obtain
\begin{align}
f(\chi) \approx A\exp\left[\frac{\pi}{2\bar{H}} - \chi\right] & \quad
\mbox {for} \quad 1 \ll \chi \ll 1/\bar{H}
\label{smallchifree}\, ,\;\mbox{and} \\
f(\chi) \approx
\frac{A}{\sqrt{\chi\bar{H}}}\;\exp\left[\frac{i}{2}\chi^2\bar{H}\right]
& \quad \mbox{for} \quad \chi \gg 1/\bar{H} \,.
\label{largechifree} 
\end{align} 
As expected, the small $\chi$ behavior of \eqref{smallchifree} is of
exactly the correct form to fit the large $\chi$ asymptotic behavior
of \eqref{eq:nonlinearsoln}.  Matching to this result sets $A$ to
$\sqrt{\frac{8}{3}}\exp(-\frac{\pi}{2\bar{H}})$ and hence fixes
the coefficient of the outgoing wave \eqref{largechifree}.

Returning to our original variables and putting together the pieces,
we find the following expressions for the oscillon
\begin{align}
\phi(X, T) \approx \epsilon\sqrt{\frac{8}{3g}} 
\cos\left( m T \sqrt{1-\epsilon^2}\right)
\mathrm{sech} \left( m X \epsilon \right)  &\quad \mbox{for} \quad |X|
\ll \epsilon/H \, ,\;\mbox{and} \\
\phi(X, T) \approx 2\epsilon\sqrt{\frac{8\epsilon}{3H|X|g}}
e^{-\frac{\pi m \epsilon^2}{2 H}}\,
\cos\left(mT\sqrt{1-\epsilon^2} - \frac{1}{2}mX^2H\right)
&\quad \mbox{for} \quad \epsilon/H \ll |X| <1/H \,.
\label{eq:farfieldfull} 
\end{align}
From here it is easy to compute the relevant components of the stress
tensor $\mathcal{T}_{\mu\nu}$ and find the rate of energy flux.  If we
take the energy stored in a region $R$ to be $E = -\int_R dX
\mathcal{T}^T_{\enspace\;T}$, then using the conservation of the
stress tensor and \eqref{eq:farfieldfull} we obtain
\begin{equation}
\frac{d}{dT}E = \mathcal{T}^X_{\enspace\;T}|_{\partial R} \approx
-(1-X^2H^2)\frac{32}{3g}m^2\epsilon^3
\exp\left[-\frac{\pi m\epsilon^2}{H}\right].
\end{equation}
where the region $R$ is taken to be a symmetric interval $R = [-X,X]$
and $X$ is far enough from the origin that \eqref{eq:farfieldfull}
holds.  Note that to leading order all $X$ dependence comes from the
curvature of the metric; if we restrict attention to $\epsilon/H \ll X
\ll 1/H$ (i.e. we consider only a small neighborhood of the oscillon)
then space looks almost flat and we obtain
\begin{equation}
\frac{d}{dT}E_\mathrm{osc} \approx - \frac{32}{3g}m^2\epsilon^3
\exp\left[-\frac{\pi m\epsilon^2}{H}\right] \,.
\end{equation}
This is our main result.  While an oscillon can live forever on a flat
background (at least as a formal perturbation series), this is no
longer the case in a de Sitter universe; instead it is forced to
radiate energy away, albeit through a mode that is exponentially
suppressed.
  
\section{Thermal initial conditions}
We would like to start our simulations using initial
conditions mimicking those of the interacting field theory defined by
the Lagrangian (\ref{lag}) at nonzero temperature $T$.  However,
constructing this equilibrium is quite difficult; although we are
essentially interested in classical physics,  a classical treatment of
this field theory at nonzero temperature suffers from the Jeans
paradox.  In a quantum treatment we avoid this problem, but we are
still unable to systematically take into account the nonlinear terms
in the Lagrangian: as is shown below, we will be interested in
temperatures $T \gtrsim m/g$, which is precisely the regime
where finite temperature perturbation theory fails.

We shall thus take a different approach and generate our initial conditions
to simulate thermal states of the \emph{free} massive scalar field.  We
note that for the reasons stated above these quasi-thermal initial
conditions are probably quite far from the true thermal equilibrium of
the full interacting theory; hence the parameters $T$ and $\hbar$
should be thought of more as measures of the amplitude and width 
of our distribution in momentum space than as directly physically
relevant quantities.  We show, however that provided $T$ is
sufficiently high, all other numerically feasible variations of these
parameters produce oscillons in copious numbers, leading us to believe
that our results are independent of any particular details of the
initial conditions.

To construct these conditions, we return to comoving
coordinates.  Since our real interest is in a numerical simulation we
impose both infrared and ultraviolet cutoffs, placing the system in a
box of comoving size $L$ and on a regular lattice with spacing $\Delta
x$.  We replace the spatial derivatives by finite differences (see the
Numerical Simulation section) and label the free field's normal modes by
$k_n=2\pi n/L$, where 
$n=-N/2+1\ldots N/2$ and $N=L/\Delta x$ is the number
of lattice points. Finally we take the scale factor
at this time to be $a_0$.

On this lattice each free mode is described by a harmonic oscillator
with frequency
\begin{equation}
\omega_n = \sqrt{\left( \frac{2\sin \frac{k_n \Delta x}{2}}{a_0\Delta x}
\right)^2 + m^2} \,.
\end{equation}
The initial conditions for
$\phi$ are then given by
\begin{eqnarray}
\phi(x_j,t=0) &=& \sum_{n=-N/2 + 1}^{N/2} \sqrt{\frac{\hbar}{2 L a_0\omega_n}}
\left[ \alpha_n e^{i k_n x_k} + \alpha_n^\ast e^{-i k_n x_k} \right] \,,
\;\mbox{and}\cr
\dot \phi(x_j,t=0) &=& \sum_{n=-N/2 + 1}^{N/2} 
\frac{1}{i} \sqrt{\frac{\hbar \omega_n}{2 L a_0}}
\left[ \alpha_n e^{i k_n x_k} - \alpha_n^\ast e^{-i k_n x_k} \right] \,,
\end{eqnarray}
where $\alpha_n$ is a random complex variable with phase distributed
uniformly on $[0,2\pi)$ and magnitude drawn from a Gaussian
distribution such that
\begin{equation}
\langle|\alpha_n|^2 \rangle = 
\frac{1}{2}\left(\coth \frac{\hbar\omega_n}{2T} - 1\right)\,.
\end{equation}
This is the usual amplitude distribution for a quantum harmonic
oscillator \cite{LL} with the zero-point motion subtracted.
On average, these initial conditions assign energy $T$ to modes with
$\hbar\omega_n \lesssim T$, in agreement with equipartition.
The energy per mode goes rapidly to zero for $\hbar\omega_n \gtrsim T$,
giving a total energy density that scales like $T^2/\hbar$ for $T\gg
m\hbar$, as usual for blackbody radiation in one dimension.  Oscillons
will form from the energy density in modes with wavelengths of order
$1/m$, which scales like $mT$.  For oscillons to form, this
energy density must be at least of order $m^2/g$, so that the
fields will have amplitude $1/\sqrt{g}$ and the
nonlinear interaction terms can balance the dispersive
gradient terms.  Therefore we will need an initial temperature of
$T \gtrsim m/g$ to form oscillons.

We have simply subtracted off the zero-point quantum fluctuations
in the field. Although it is well-known that these fluctuations have
significant consequences for the evolution of a classical field in an
expanding background \cite{Guth:1985ya}, these effects are most
important at very long length scales, of order $1/H$.  For
small $g$, the oscillon solution comprises many fundamental
excitations of the field at the length scales of order $1/m$
that are relevant to oscillon formation and stability.
Thus the quantum effects we are neglecting should not change our
results significantly. Alternatively, this quantum prescription can
simply be  thought of as describing a classical equilibrium with
short-distance cutoff $T/\hbar$. (A method to eliminate such
cutoff dependence of classical simulations is discussed in
\cite{Borrill:1996uq}).

As the universe expands, it cools and loses energy according to
\begin{equation}
\frac{dE}{dt} = -\int p(x,t) \dot a(t) dx \,,
\label{eqn:pdV}
\end{equation}
where the pressure density is 
\begin{equation}
p(x,t) = \frac{1}{2}\left[\dot \phi^2 + 
\frac{1}{a(t)^2}(\phi')^2 - m^2 \left(\phi^2
- \frac{g}{2} \phi^4 + \frac{g^2}{3} \phi^6\right)\right] \,.
\label{pden}
\end{equation}
In equilibrium at temperatures much greater than $\hbar m$, if we
neglect the interaction terms, the system looks like massless radiation,
with pressure density approximately equal to its energy density.  In
equilibrium at low temperatures, the field is slowly varying in
space, with only small values of $k$ excited, and of small amplitude.
In that case the gradient and nonlinear terms are negligible and $\dot
\phi^2 \approx m^2 \phi$, so the pressure goes to zero and the system
behaves like pressureless dust.

\section{Numerical Simulation}

We discretize $x$ at the level of the Lagrangian in Eq.~(\ref{lag}),
working in natural units where $m=1$.  For the space derivatives we
use ordinary first-order differences,
\begin{equation}
\phi'_n(t) = \frac{\phi_{n+1}(t) - \phi_{n}(t)}{\Delta x} \,,
\end{equation}
where $\phi_n$ refers to the value of $\phi$ at lattice point $n$.
We work on a regular lattice with spacing $\Delta x$ and impose
periodic boundary conditions.  Varying this Lagrangian yields lattice
equations of motion with second-order space derivatives,
\begin{equation}
\ddot \phi_n(t) + \frac{\dot a(t)}{a(t)} \dot \phi_n(t) = 
\frac{\phi_{n+1}(t) + \phi_{n-1}(t) - 2 \phi_{n}(t)}{a(t)^2 \Delta x^2}
- \phi_n(t) + g \phi_n(t)^3 - g^2 \phi_n(t)^5 \,.
\end{equation}

We then express this second order equation as a set of coupled first order
differential equations and step forward in time using a standard
fourth-order Runge-Kutta integrator (e.g., see \cite{numrec}).  The
Courant condition requires that we maintain $\Delta t < a(t) \Delta x$
for numeric stability.  Of course this allows us the possibility of
rescaling $\Delta t$ as the simulation runs; this rescaling stops when
we reach a maximum value of $\Delta t = 0.01$.

As the universe expands, the oscillons maintain a fixed size in
physical units. On the other hand, our lattice expands with the
universe.  Thus we add new lattice points whenever the lattice spacing
exceeds some fixed size $\Delta x_{\mathrm{max}}$ in physical
units.  This is accomplished by refining the lattice, doubling the
total number of lattice points and bringing the lattice spacing back
to $\Delta x_{\mathrm{max}}/2$ in physical units.  We assign values to
the field for the new intermediate lattice points by linear interpolation. 

We performed several checks on our numerics.  We would like to 
take $\Delta t$ sufficiently small that given a set of initial data
any further reduction of $\Delta t$ does not significantly change the
final configuration after a run.  In the strongly nonlinear regime in
which we work, this turns out to be a technically very difficult goal
to achieve, with different timesteps often resulting in significantly
different final configurations. In fact, despite our best efforts the
upper end of Fig. \ref{fig2}, where $T > 6$, suffers from this
problem, so we cannot guarantee the validity of this region of the
plot, although the remainder of the plots we present satisfy all of
our tests. We also verify that all of our simulations maintain energy
conservation (as given by Eq. (\ref{eqn:pdV})) to better than one part
in $10^3$, and that $\Delta x_{\mathrm{max}}$ is small enough that
further reduction does not significantly alter the final
configuration obtained after a run.

For Figure \ref{fig1} we use an initial lattice spacing $\Delta x_0 =
0.0016$ and an initial timestep of $\Delta t_0 = 0.0002$. For Figure
\ref{fig2} we use $\Delta x_0 = 0.05/T$ and $\Delta t_0 = 0.0001$,
where $T$ is the temperature of the run.  For both we use $\Delta
x_{\mathrm{max}} = 1$.

\section{Results}

\begin{figure}[htbp]
\includegraphics[width=0.92\linewidth]{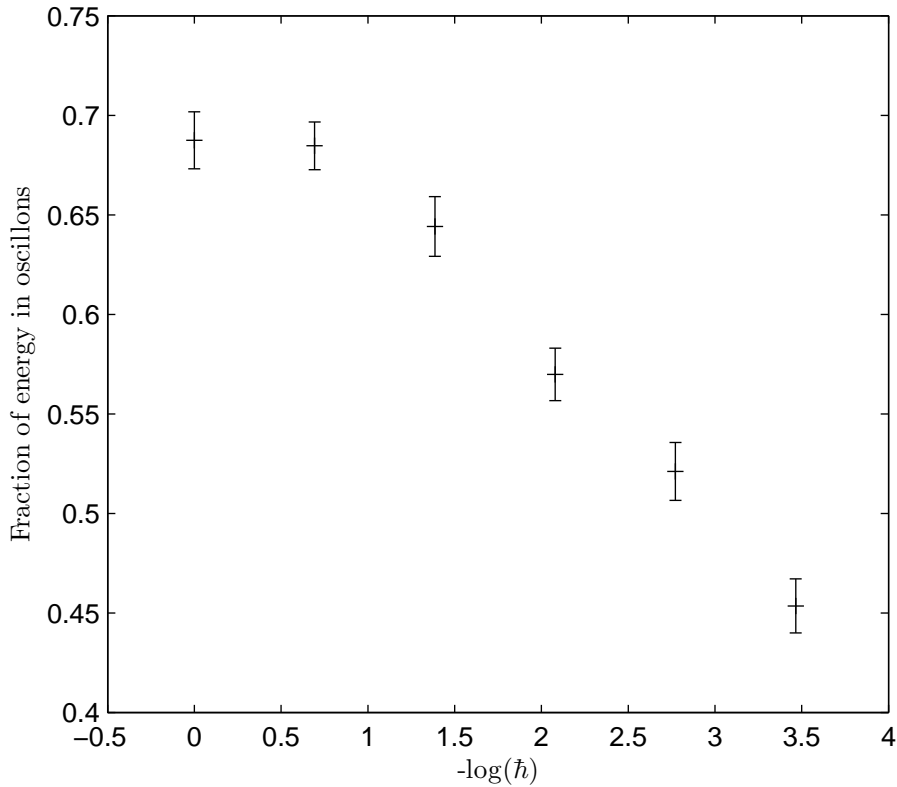}
\caption{
Fraction of energy in oscillons, as a function of $\log
(\frac{1}{\hbar})$.  Simulation parameters are $T = 2$, $g = 1$,
initial universe size $L_0 = 20$, expansion factor = $2000$, Hubble
constant $H = 0.02$.
}
\label{fig1}
\end{figure}

\begin{figure}[htbp]
\includegraphics[width=0.95\linewidth]{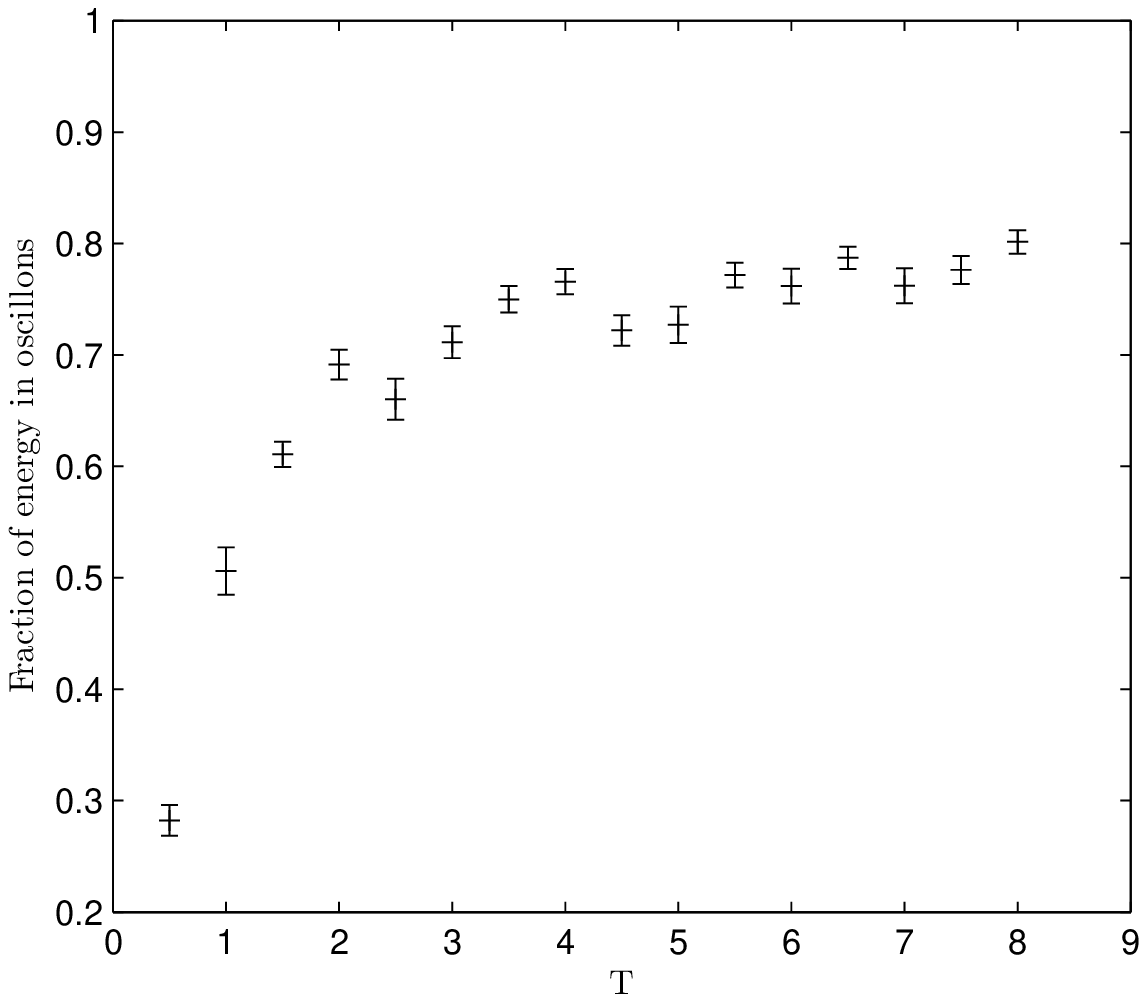}
\caption{
Fraction of energy in oscillons, as a function of initial temperature.
Simulation parameters are $\hbar = 0.5$, $g = 1$, initial
universe size $L_0 = 40/T$, expansion factor $ = 40000/L_0$, Hubble constant
$H = 0.02$.}
\label{fig2}
\end{figure}

We simulate the scalar field in an expanding universe varying the
initial temperature $T$ and the value of $\hbar$ (which determines the
effective coupling $g\hbar$).  We are interested in determining what
fraction of energy in the universe ends up in oscillons; we estimate
this as the integral of the  energy density over regions of space at
which the energy density is more than five times the average energy
density, divided by the total energy. As shown in Fig.~\ref{fig3},
oscillons stand so much higher than the background fluctuations that
this is a good measure.  For each run we expand until this quantity
reaches a constant value, indicating that the system has stabilized.
We then plot this quantity as a function of the parameters in the
initial conditions in Fig.~\ref{fig1} and Fig.~\ref{fig2}.

Fig.~\ref{fig3} shows a snapshot of the energy density as a
function of position for a typical run.  We see that the oscillons are
very clearly defined, with energy densities far above the background from
ordinary field fluctuations.  At these late times, the ordinary
fluctuations are of small amplitude, and have been redshifted to large
spatial wavelengths; they have $k\approx 0$ and $\omega \approx m$.  From
Eq.~(\ref{pden}), we see that the pressure of such fluctuations
vanishes:  the $\dot \phi^2$ and $m^2 \phi^2$ terms cancel, and the
gradient and nonlinear terms can be neglected because the ordinary
fluctuations are slowly varying in space and have small amplitude.  So
the total energy in ordinary fluctuations remains constant,
corresponding to a pressureless dust of ordinary particles of mass
$\hbar m$ at rest.  As the universe continues to expand, the energy
density in these fluctuations decreases, since it scales inversely
with the volume of the universe to keep the total constant.  The
oscillons, meanwhile, maintain fixed physical size, energy density,
and total energy --- and therefore zero pressure --- but are
brought to rest relative to the expanding background.  As a result, 
they also appear as a pressureless dust at late times.

In Fig.~\ref{fig1} we vary $\hbar$ in the initial conditions and 
examine the variation of the fraction of energy in the universe in
oscillons.  Each point on the graph is an average over many runs, with
error bars indicating the standard deviation.  (Because our initial
conditions are random individual representatives drawn from the
initial thermal distribution, our results can vary from one run to the
next.)  As we move to the right on the logarithmic scale on the
horizontal axis, $\hbar$ becomes smaller and the quantum effects we
have neglected are less important.  We see that the fraction of energy
in oscillons remains substantial, decreasing only gradually as $\hbar$
decreases to small values. 

Fig.~\ref{fig2} shows the fraction of energy in oscillons for a
range of initial temperatures.  For $T$ large enough, we see that the
result saturates, so that higher initial temperatures no longer affect
the final result.  In the saturated regime, the system is just
undergoing ordinary cooling, remaining at equilibrium as the
temperature begins to decrease.  It is only once the system cools
below $T \approx m/g$ that the oscillons begin to emerge.  So while we are
always imagining our system starts out at very high temperatures
compared to the energy scales relevant to oscillon formation, in
practice we only need to start our simulations at temperatures
just above the saturation point.

\section{Discussion and Conclusions}
We have considered a scalar field theory that supports stable
oscillons.  We place this theory in an expanding background and show
that the oscillons are no longer completely stable but instead lose
energy at a rate that is exponentially small in the size of the
horizon.  We also find numerically that quasi-thermal initial
conditions result in the eventual formation of oscillons in large
numbers.  Though the oscillons are large, coherent objects, they
nevertheless form easily from a random superposition of momentum
modes, suggesting that in some sense they are attractors in the
solution space of the equations of motion.  In our model they capture
a significant portion of the energy of the universe, on the order of
50\%.  Though this paper deals with a one-dimensional case, we have
seen qualitatively similar results in two dimensions. If this
phenomenon persists in realistic models (and in a realistic number of
dimensions), oscillons may have cosmological consequences, as
discussed in Ref. \cite{Gleiserinfl}.
  
\section{Acknowledgments}

We thank N.\ Alidoust, A.\ Scardicchio, and
R.\ Stowell for assistance and M.\ Gleiser for helpful discussions.
E.\ F.\ and A.\ H.\ G.\ were supported in part by funds provided by
the U.\ S.\ Department of Energy (D.\ O.\ E.) under cooperative research
agreement DE-FC02-94ER40818.  N.\ G.\ and N.\ S.\ were supported by
National Science Foundation (NSF) grant PHY-0555338, by a Cottrell
College Science Award from Research Corporation, and by Middlebury
College. N.\ I.\ was supported in part by National Science Foundation
(NSF) Graduate Fellowship 2006036498.

\end{document}